\documentclass[twocolumn,amsfonts,showpacs,superscriptaddress]{revtex4-1}
\usepackage[dvipdfmx]{graphicx}
\usepackage{amssymb,amsmath,amsthm,booktabs,mathtools}
\usepackage{bbm}
\usepackage{bm}
\usepackage{color}
\usepackage[colorlinks,bookmarks=false,citecolor=blue,linkcolor=red,urlcolor=blue]{hyperref}
\usepackage{tikz}
\usepackage{pgfplots}
\usepackage{enumitem}

\def\tit#1{}

\usepackage{color}

\usepackage{amsmath}	
\begin{document}

\title{Quantum Generalized Hydrodynamics}

\author{Paola Ruggiero}
\affiliation
{SISSA and INFN, Via Bonomea 265, 34136 Trieste, Italy}
\author{Pasquale Calabrese}
\affiliation
{SISSA and INFN, Via Bonomea 265, 34136 Trieste, Italy}
\affiliation
{International Centre for Theoretical Physics (ICTP), Strada Costiera 11, 34151 Trieste, Italy}
\author{Benjamin Doyon}
\affiliation
{Department of Mathematics, King’s College London, Strand WC2R 2LS, UK}
\author{J\'er\^ome Dubail}
\affiliation
{Laboratoire de Physique et Chimie Th\'eoriques, CNRS, UMR 7019, Universit\'e de Lorraine, 54506 Vandoeuvre-les-Nancy, France
}

\begin{abstract}
Physical systems made of many interacting quantum particles can often be described by Euler hydrodynamic equations in the limit of long wavelengths and low frequencies. Recently such a classical hydrodynamic framework, now dubbed {\it Generalized Hydrodynamics} (GHD), was found for quantum integrable models in one spatial dimension. Despite its great predictive power, GHD, like any Euler hydrodynamic equation, misses important quantum effects, such as quantum fluctuations leading to non-zero equal-time correlations between fluid cells at different positions. Focusing on the one-dimensional gas of bosons with delta repulsion, and on states of zero entropy, for which quantum fluctuations are larger, we reconstruct such quantum effects by quantizing GHD. The resulting theory of {\it quantum GHD} can be viewed as a multi-component Luttinger liquid theory, with a small set of effective parameters that are fixed by the Thermodynamic Bethe Ansatz. It describes quantum fluctuations of truly nonequilibrium systems where conventional Luttinger liquid theory fails.
\end{abstract}

\maketitle

The behavior of fluids at very low temperatures is usually peculiar as the quantum nature of their constituents dominates over thermal fluctuations. To describe collective quantum effects, it is customary to start from classical hydrodynamic equations, and to quantize them. This path was taken by Landau in 1941~\cite{landau1941theory} in his development of the theory of superfluid helium \cite{khalatnikov2018introduction,putterman1974superfluid}. Since then, similar approaches have been developed for various other quantum liquids~\cite{nozieres2018theory,leggett2006quantum}, including for instance Bose-Einstein condensates where quantum fluctuations are captured by the Bogoliubov theory~\cite{bogolyubov1947theory,pitaevskii2016bose,mora2003extension}, quantum Hall liquids~\cite{wiegmann2013hydrodynamics,wiegmann2014anomalous,wiegmann2019quantization}, 
or one-dimensional (1d) quantum fluids described by Luttinger liquid theory~\cite{haldane1981luttinger,giamarchi2003quantum,cazalilla2004bosonizing}.

The purpose of this Letter is to apply a similar program to the classical hydrodynamics of one-dimensional quantum integrable models introduced in 2016~\cite{bertini2016transport,castro2016emergent}, now dubbed {\it Generalized Hydrodynamics} (GHD). At equilibrium, the Luttinger
liquid theory ---which is the quantized hydrodynamic theory
of 1d fluids~\cite{abanov2006hydrodynamics} with few conserved quantities such
as charge, magnetization, energy, momentum--- is enough to capture
quantum fluctuations of 1d gapless integrable models~\cite{haldane1981luttinger,giamarchi2003quantum,cazalilla2004bosonizing}. However,
when dealing with true out-of-equilibrium situations, like the quantum Newton's cradle setup~\cite{kinoshita2006quantum} ---whose hydrodynamics description must keep track of all higher conservations laws~\cite{rigol2007relaxation}, and is provided by GHD~\cite{caux2019hydrodynamics}---, quantum fluctuations must be given by a more general quantum hydrodynamics theory, obtainable by quantizing GHD.
Here our goal is to identify that theory.

\begin{figure}[ht]
	\begin{tikzpicture}[scale=0.5]
		\draw (0,0) node{\includegraphics[width=0.48\textwidth]{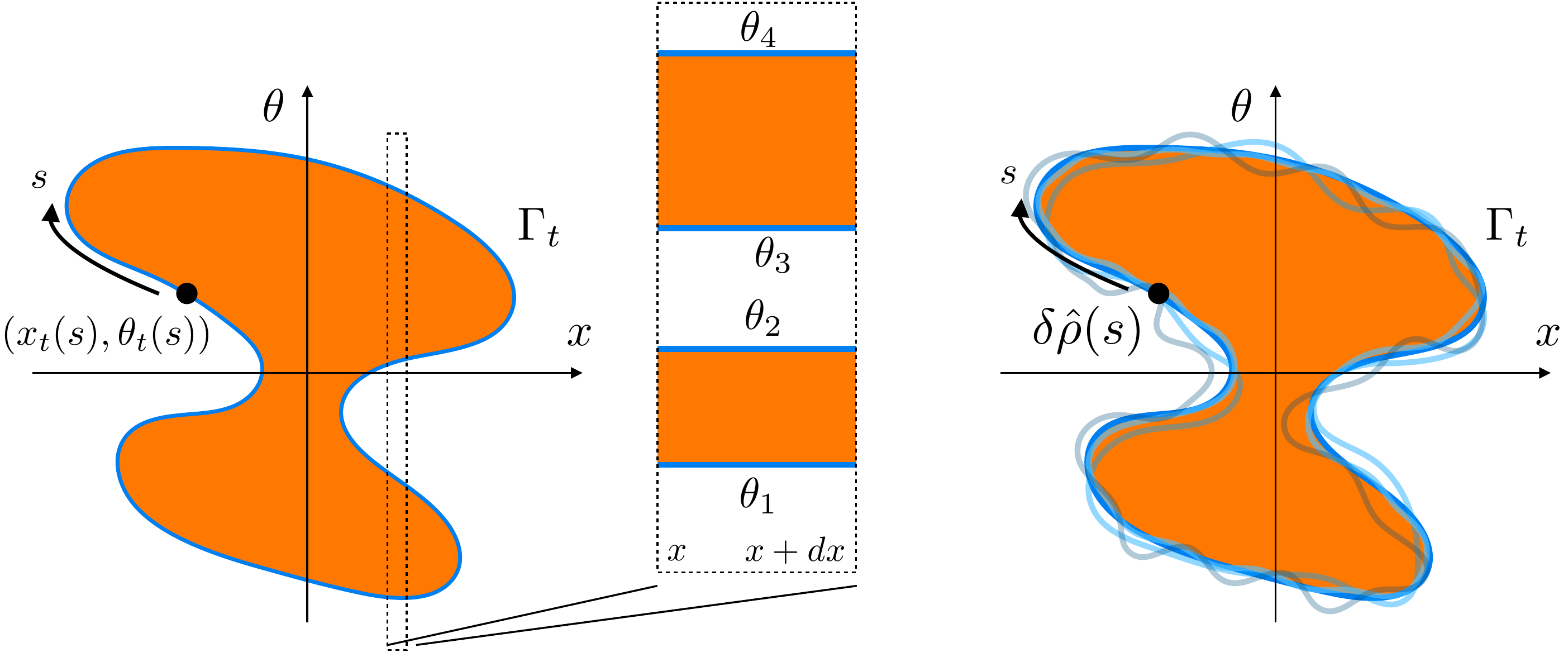}};
		\draw (-8,-3) node{$a)$};
		\draw (3,-3) node{$b)$};
	\end{tikzpicture}
	\vspace{-0.6cm}
	\caption{(a) Zero-entropy GHD describes the motion of the Fermi contour $\Gamma_t$, parametrized as in Eq. (\ref{eq:params}), which separates the regions in phase space where the Fermi factor $n(x,\theta)$ is one (orange) or zero (white) at a given time $t$. In any small interval $[x,x+ d x]$ the fluid is in a state called {\it split Fermi sea} \cite{fokkema2014split,eliens2016general,vlijm2016correlations,eliens2017quantum} labeled by Fermi rapidities $\theta_1 < \theta_2 < \dots < \theta_{2q}$; the number of fluid components $q$ is a piecewise constant function of $x$ and $t$. (b) In this Letter the contour $\Gamma_t$ is allowed to have quantum fluctuations around the classical solution to the zero-entropy GHD equations (\ref{eq:GHD}a-b). The quantum fluctuations are captured by a chiral boson with density $\delta \hat{\rho} (s)$ living along the contour.
	}
	\label{fig:Fermi}
\end{figure}

Our starting point is GHD, which, on the technical side, relies on the formalism of the Thermodynamic Bethe Ansatz~\cite{yang1969thermodynamics,takahashi2005thermodynamics}. Thermodynamically large integrable systems are described by densities of different species of quasiparticles. For simplicity,
in this Letter we formulate our results in a specific model: the 1d Bose gas with delta repulsion~\cite{lieb1963exact,berezin1964schrodinger,korepin1997quantum}. This model is singled out because of its experimental relevance ---it is routinely used for describing contemporary cold atom experiments in 1d~\cite{olshanii1998atomic,van2008yang,vogler2013thermodynamics,schemmer2019generalized,wilson2019observation}--- and because of its simple thermodynamics involving a {\it single} species of quasiparticles. Our approach can be straightforwardly generalized to other integrable systems with a GHD description~\cite{piroli2017transport,ilievski2017microscopic,ilievski2017ballistic,doyon2017large,doyon2017drude,bulchandani2018bethe,collura2018analytic,doyon2018soliton,cao2018incomplete,doyon2018exact,bastianello2018generalized,de2018hydrodynamic,10.21468/SciPostPhys.6.4.049,gopalakrishnan2018hydrodynamics,mazza2018energy,vu2018equations,borsi2019current,doyon2019generalised,bulchandani2019kinetic,spohn2019generalized,bastianello2019generalized,panfil2019linearized,alba2019towards,agrawal2019generalized,cubero2019generalized,bertini2019GHD,mestyan2019GHD,bertini2018GHD,bertini2019GHDb,mestyan2019GHDb}, including for instance the XXZ chain; we defer mathematical formulas for the general multi-species case to the Supplemental Material (SM)~\cite{SM}.

At the microscopic level, the 1d Bose gas with delta repulsion is defined by the Hamiltonian for $N$ bosons $H = \sum_{i=1}^N [-\frac{\hbar^2}{2} \partial_{x_i}^2 + V(x_i)] + \hbar \bar{g} \sum_{i<j} \delta(x_i-x_j) $, where $g = \hbar \bar{g} >0$ is the repulsion strength between the bosons and $V(x)$ is an external trapping potential. 
We set the mass of the bosons to $1$. 

GHD is formulated at the Euler scale, where space-time scales of observations and length scales of external potentials are simultaneously sent to infinity; at the Euler scale diffusion is absent (but subleading diffusive corrections to GHD are also known  \cite{de2018hydrodynamic,10.21468/SciPostPhys.6.4.049,gopalakrishnan2018hydrodynamics}). In the 1d Bose gas the Euler scale is equivalently expressed as the  classical limit~\cite{brun2017one,brun2018inhomogeneous,ruggiero2019conformal}
\begin{equation}
	\label{eq:limit}
	 \hbar \rightarrow 0 , \quad \; {\rm keeping} \quad \hbar N, \; V(x), \; \bar{g} \; \; {\rm fixed }.
\end{equation}
For the gas starting at {\it zero temperature}, its evolution under GHD takes a particularly simple form~\cite{doyon2017large}. Indeed, an initial zero-temperature state has zero entropy and entropy is conserved by Euler equations such as GHD. [This is generally true for Euler hydrodynamic equations away from shocks, and it is known that GHD does not admit shocks~\cite{el2005kinetic,doyon2017large,bulchandani2017classical}.] Thus the Bose fluid remains locally in a macrostate with zero entropy at all times. In the Bose gas with delta repulsion, the presence of higher conservation laws allows for a large space of macrostates with zero entropy: the {\it split Fermi seas}~\cite{fokkema2014split,eliens2016general,vlijm2016correlations,eliens2017quantum}. They can be labeled by a set of {\it Fermi rapidities} $\{\theta_a\}_{1\leq a\leq 2q}$ such that the {\it Fermi factor} $n(\theta)$ ---the number of Bethe quasiparticles with rapidity in $[\theta, \theta + d \theta]$ divided by the number of available states in that interval (see e.g. Chap. 1 in Ref.~\cite{korepin1997quantum}  for an introduction to that formalism)--- is
\begin{equation}
	n(\theta) \, = \, \left\{ \begin{array}{rcl} 
		1 & {\rm if} & \theta \in [\theta_1 ,\theta_2] \cup \dots \cup [\theta_{2q-1},\theta_{2q}] \\
		0 & & {\rm otherwise} .
		\end{array}
	 \right.
\end{equation}
The local macrostate is then assumed to be a function of position $x$ and of time $t$. The global state of the system at time $t$ is best represented by the {\it Fermi contour} $\Gamma_t$ (see Fig. \ref{fig:Fermi}), which is defined such that the Fermi factor $n(x,\theta)$ is $1$ for all points $(x,\theta)$ inside the contour, and $0$ outside. For simplicity we restrict to situations where $\Gamma_t$ is a simple closed curve, parametrized by a function $s \mapsto (x_t(s), \theta_t(s))$,
\begin{equation}
	\label{eq:params}
	\Gamma_t = \{ (x_t(s), \theta_t(s)) , s \in  \mathbb{R}/ 2\pi \mathbb{Z} \}  .
\end{equation}
According to GHD, the time evolution of the contour $\Gamma_t$ is given by the classical equation~\cite{doyon2017large}
\begin{subequations}
	\label{eq:GHD}
\begin{equation}
	\frac{d}{dt} \left( \begin{array}{c}
		x_t(s) \\
		\theta_t(s)
	\end{array} \right) \, = \,  \left( \begin{array}{c}
		v^{\rm eff} (x_t(s),\theta_t(s))  \\
		a^{\rm eff}(x_t(s),\theta_t(s))
 	\end{array} \right) ,
\end{equation}
which expresses the fact that quasiparticles inside the contour move at an effective velocity $v^{\rm eff}(x,\theta)$ and are accelerated at an effective acceleration $a^{\rm eff}(x,\theta)$, both of which depend in general on the local Hamiltonian and macrostate, hence on the Fermi points at $x$. Eq. (\ref{eq:GHD}a) is complemented by a closed formula for the effective velocity~\cite{bonnes2014light,bertini2016transport,castro2016emergent,vu2019equations} and acceleration~\cite{doyon2017note}
\begin{equation}
	v^{\rm eff}  \, = \, (\partial_\theta E)^{\rm dr} / 1^{\rm dr}   ,\quad
	a^{\rm eff}  \, = \, -(\partial_xE)^{\rm dr} / 1^{\rm dr} ,
\end{equation}
\end{subequations}
where $E(x,\theta)$ is the bare energy of a quasiparticle with respect to the local Hamiltonian, $1(\theta) = 1$, and the {\it dressing} of a function $f(\theta)$ in the local macrostate is defined by the integral equation $f^{\rm dr} (\theta) \, = \, f(\theta) +  \int \frac{d\theta'}{2\pi} \frac{d \phi (\theta-\theta')}{d \theta} n(\theta') f^{\rm dr} (\theta')$. Here $\phi(\theta-\theta') = 2 \, {\rm arctan} \left( (\theta-\theta')/\bar{g} \right)$ is the two-body scattering phase for the delta Bose gas~\cite{lieb1963exact,berezin1964schrodinger,korepin1997quantum}. In the present case, $E(x,\theta) = \theta^2/2 + V(x)$, and the effective acceleration simplifies to give Newton's second law~\cite{doyon2017note}
\begin{equation}
	\label{eq:newton}
	a^{\rm eff}= -\partial_x V(x).
\end{equation}
Notice that $\hbar$ is completely absent from the Eqs. (\ref{eq:GHD}a,b), which is consistent with our claim that zero-entropy GHD corresponds to the classical limit (\ref{eq:limit}) in the microscopic model.

\begin{figure*}[t]
\centering
\includegraphics[width = 0.99\textwidth]{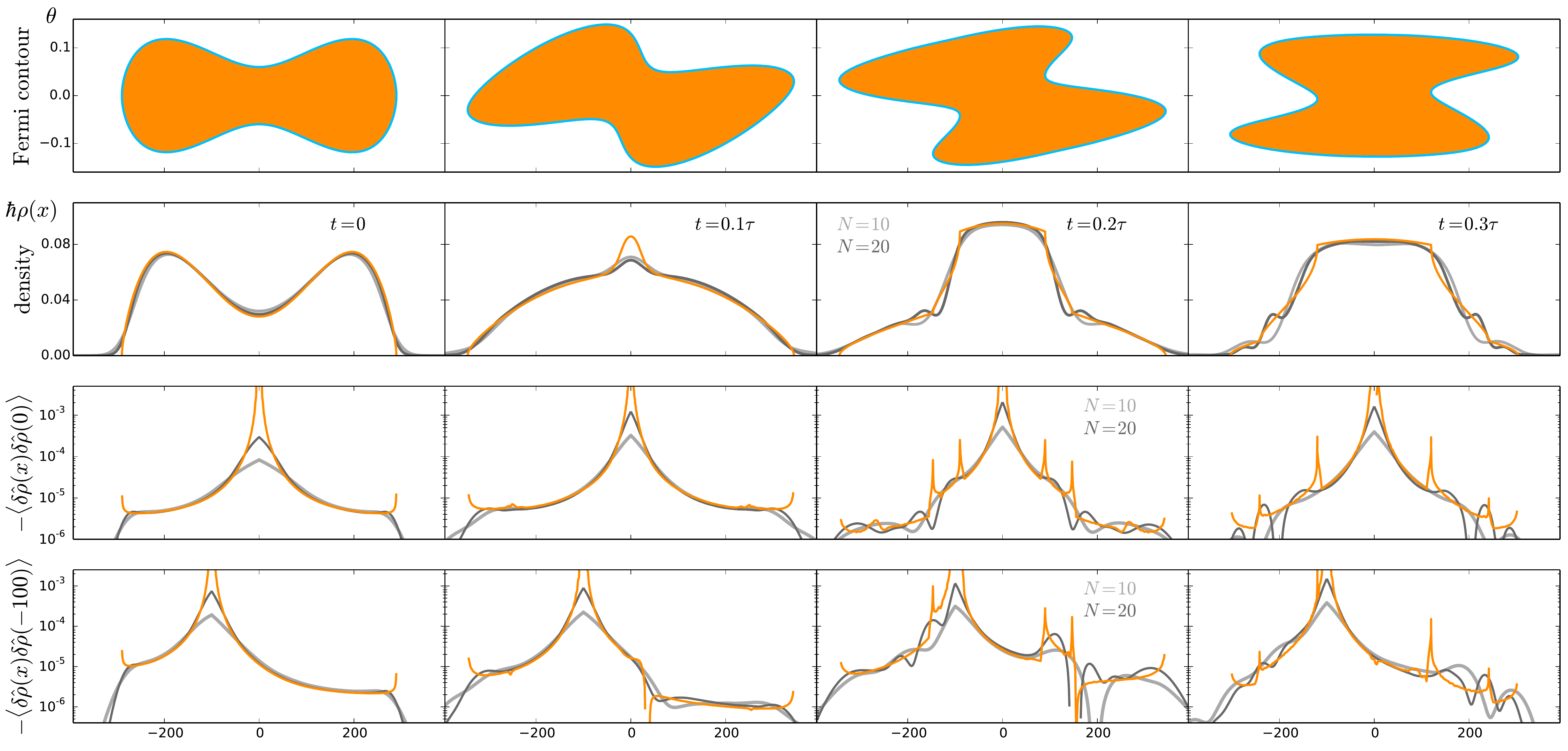}
\caption{Quantum quench from double to single well in the 1d Bose gas with delta repulsion. We compare the predictions of GHD and QGHD (orange curves) to time-dependent DMRG simulation for N=10 (light gray) and N=20 particles (dark gray). First row: Fermi contour evolved with GHD. Second row: density profile predicted by GHD, compared with DMRG. Third and fourth row: connected density-density correlator $\left< \hat{\rho} (x) \hat{\rho} (x_0)\right>$ predicted by QGHD and compared with DMRG, for two different positions $x_0$. Each row shows the corresponding quantity as a function of the spatial coordinate $x$, at different times, expressed as a fraction of the period $\tau$ (from $t=0$ in the first column to $t=0.3 \tau$ in the last). 
For the DMRG simulation we work with particles on a lattice at very low density. The parameters are: repulsion strength $\bar{g} = 0.1$; $L=800$ lattice sites; number of particles $N=10, 20$; $\hbar = 30/N$; pre-quench potential $V_0(x)= (x/L)^4 - 0.12 (x/L)^2$; post-quench potential $V(x) = \omega^2 x^2 /2$ with $\omega=0.3/L$ (and period $\tau= 2\pi/\omega$). The dimensionless Lieb parameter $\gamma = \frac{\bar{g}}{\hbar \rho}$ is of order $1$, so we are far from both the Gross-Pitaevski limit and the Tonks-Girardeau limit.}
\label{movie}
\end{figure*}

\vspace{0.1cm} \noindent {\bf\em Goal of this Letter.}\; Because it is a classical hydrodynamic description, GHD misses certain quantum effects, such as quantum entanglement or correlations between the different parts of the fluid at a given time. Such effects appear as subleading orders in an expansion at small $\hbar$ in the limit (\ref{eq:limit}). Here we initiate the development of a theory of {\it quantum fluctuations around GHD}. Analogously to Bogoliubov theory~\cite{bogolyubov1947theory,pitaevskii2016bose,mora2003extension}, our strategy is to start from linear sound waves propagating on top of a background configuration $(x_t(s), \theta_t(s))$ which solves the GHD equation (Fig.~\ref{fig:Fermi}), and then find a way to quantize those. We find that the resulting theory takes the form of a {\it time-dependent, spatially inhomogeneous, multi-component Luttinger liquid}, which generalizes the effective theory of (homogeneous, time-independent) split Fermi seas developed recently by Eli\"ens and Caux~\cite{eliens2016general}, see also Refs.~\cite{fokkema2014split,vlijm2016correlations,eliens2017quantum}. It also generalizes the theory of inhomogeneous Luttinger liquids (see e.g. Refs.~\cite{gangardt2003stability,abanov2006hydrodynamics,dubail2017conformal,brun2017one,brun2018inhomogeneous,ruggiero2019conformal,cazalilla2004bosonizing}) to truly out-of-equilibrium situations, like the situation depicted in  Fig.~\ref{movie} (see the discussion below).


\vspace{0.1cm} \noindent {\bf\em Sound waves in zero-entropy GHD.}\; Linearly propagating waves are consequences of the conservation laws of hydrodynamics. By fluctuation-dissipation, they are subject to correlations due to microscopic fluctuations. Thus, our first task is to find conserved fluid modes and their linear-response evolution. Let us parametrise locally the contour $\Gamma_t$ by the Fermi points $\theta_a(x,t)$ ($1 \leq a \leq 2q$). Fluctuations can be expressed as deformations of the contour $\theta_a(x,t)\rightarrow \theta_a(x,t)+\delta \theta_a(x,t)$. Plugging this into Eqs. (\ref{eq:GHD}a,b), one would arrive at an evolution equation for $\delta \theta_a (x,t)$, describing the propagation of sound waves on top of the background solution $(x_t (s) , \theta_t(s) )$. These, however, do not take the form of conservation equations.

Instead, we consider the momentum and energy of an excitation
~\cite{korepin1997quantum}, $p(\theta) = \theta + \int \frac{d\theta'}{2\pi} \phi(\theta-\theta') n(\theta')1^{\rm dr}(\theta')$ and $\epsilon(\theta) = E(x,\theta) + \int \frac{d\theta'}{2\pi} \phi(\theta-\theta')n(\theta')1^{\rm dr}(\theta')v^{\rm eff}(\theta')$, respectively. The dispersion relation of such an excitation is the effective velocity, $\partial_\theta \epsilon/ \partial_\theta p = v^{\rm eff}$. In the theory of GHD~\cite{bertini2016transport,castro2016emergent}, any conserved charge of the form $q = \int \frac{d \theta'}{2\pi} f(\theta') n(\theta')1^{\rm dr} (\theta')$, which counts $f(\theta')$ for every quasiparticle $\theta'$, satisfies a continuity equation with the current $j =\int \frac{d \theta'}{2\pi} f(\theta') n(\theta') 1^{\rm dr} (\theta') v^{\rm eff}(\theta')$. In an external potential, the continuity equation includes the effective acceleration (\ref{eq:newton}), see Ref.~\cite{doyon2017note}: $\partial_t q + \partial_x j =  \int \frac{d \theta'}{2\pi} f'(\theta') n(\theta') 1^{\rm dr} (\theta') a^{\rm eff}(\theta')$. We observe that the second terms in the expressions of $p(\theta), \epsilon(\theta)$ are precisely of the form $q, j$ (with $f(\theta') = \phi(\theta-\theta')$ \footnote{This is the ``scattering charge'', counting the total phase accumulated by the $\theta$-quasiparticle.}) therefore the conservation equation with force term holds,
\begin{equation}\label{eq:k}
	\partial_t p + \partial_x \epsilon + a^{\rm eff} 1^{\rm dr} = 0.
\end{equation}
Excitations in a zero-entropy state occur at the Fermi points. Combining the dispersion relation, \eqref{eq:k} and \eqref{eq:GHD}, one finds an {\em exact conservation law} for their momentum $p_a = p(\theta_a)$  and energy $\epsilon_a = \epsilon(\theta_a)$ (see Supplemental Material (SM) for detailed derivation~\cite{SM}):
\begin{equation}\label{eq:ka}
	\partial_t p_a + \partial_x \epsilon_a = 0.
\end{equation}
Then the small fluctuations obey, at first order,
\begin{equation}\label{eq:deltaka}
	\partial_t \delta p_a + \sum_b\partial_x[\mathsf A_a^{~b} \delta p_b] = 0,
\end{equation}
where $ \mathsf A_a^{~b} = \partial \epsilon_a/\partial p_b$ is the {\it flux Jacobian}. This is the propagation equation we were looking for: it is an equation for linear sound waves which takes the form of a conservation equation.

\vspace{0.1cm} \noindent {\bf\em Quantization of sound waves.}\; The conserved modes $\delta p_a$ can now be given quantum fluctuations, $\delta p_a\rightarrow \delta\hat p_a$. In quantized fluid theory one assumes that there is a classical hydrodynamic action $S = S(\{p_a\})$, whose minimum gives rise to the fluid equation, and which provides the quantum fluctuations and long-range correlations simply by quadratic expansion:
\begin{equation}
	e^{iS} \approx e^{iS_{\rm classical} + i\sum_{ab} S^{(2)}_{ab} \delta p_a \delta p_{b} }.
\end{equation}
Passing to the Hamiltonian formalism, there must be a symplectic structure and a Hamiltonian, quadratic in hydrodynamic wave operators $\delta \hat k_a=\delta \hat p_a/\hbar$, which reproduces \eqref{eq:deltaka}.

To identify those, consider the measure $d p = 1^{\rm dr} d\theta$, which takes into account the density of allowed states $1^{\rm dr}$ \cite{korepin1997quantum}, and the phase-space volume form it induces, $dx \wedge dp = 1^{\rm dr}\, dx \wedge d\theta$. This volume form is preserved by GHD~\cite{doyon2018geometric}. Therefore, the fluctuations at zero entropy are fluctuations of an incompressible region in the $(x,p)$ plane. A first consequence is that small volume variations $d p^a = \sigma_a dp_a$, where $\sigma_a = (-1)^a$ is the chirality of the volume boundary, are thermodynamic potentials, leading to an Onsager reciprocity relation (see SM \cite{SM})
\begin{equation}\label{eq:Asym}
	\mathsf A^{ab} = \mathsf A^{ba} \quad
	(\mathsf A^{ab} = \partial \epsilon_a / \partial k^b
	=\sigma_b \mathsf A_a^{~b}).
\end{equation}
That is, the diagonal matrix $\sigma = {\rm diag}( \{ \sigma_a \}_{1\leq a \leq 2q})$ gives a symplectic structure under which the flux Jacobian is symmetric. Second, the problem of quantizing fluctuations of incompressible regions is well known in the literature on the quantum Hall effect~\cite{wen1990chiral,wen1992theory,iso1992fermions,cappelli1993infinite}. Parameterizing the boundary of that region as $(x(s), p(s))$ and introducing a density operator which measures the excess number of occupied states around $(x(s), p(s))$, $\delta \hat{\rho} (s) = \frac{1}{2\pi \hbar} \frac{dx}{ds}  \delta \hat{p} (x)$, the commutation relation is the one of a chiral U(1) current algebra,
\begin{subequations}
\begin{equation}
	\label{eq:comm_rho}
	\left[ \delta \hat{\rho} (s) ,   \delta \hat{\rho} (s') \right] \, = \, \frac{1}{2\pi i}   \, \delta' (s-s') .
\end{equation}
Equivalently, with the local parameterization $\delta \hat{p}_a(x)$,
\begin{equation}
	\label{eq:comm_p}
	\left[ \delta \hat{p}_a (x) ,   \delta \hat{p}_b (y) \right] \, = \, -i \sigma_a  2\pi \hbar^2   \delta_{ab}  \, \delta' (x-y) .
\end{equation}
\end{subequations}
Using this symplectic structure, the Hamiltonian generating \eqref{eq:deltaka} can be taken as
\begin{equation}
	\label{eq:hamiltonian}
	\hat H[{\Gamma_t}] \,=\,  \frac{1}{4 \pi \hbar} \int dx  \sum_{a,b}   \delta \hat p_a(x) \mathsf A^{ab} \delta \hat p_b(x)  .
\end{equation}
Indeed, together with the commutation relation (\ref{eq:comm_p}), the Heisenberg equation
\begin{equation}
	\frac{d}{dt} \delta \hat p_a (x) = \frac{i}{\hbar} [\hat H[\Gamma_t], \delta \hat p_a (x) ] 
\end{equation}
reproduces the equation for sound waves (\ref{eq:deltaka}).

The dependence of $\hat H[\Gamma_t]$ on $\Gamma_t$ is via that of $\mathsf A^{ab}$ on the Fermi points $\{\theta_c(x,t)\}$. The contour-dependent Hamiltonian (\ref{eq:hamiltonian}) is the most important result of this Letter, and we refer to it as the {\it QGHD Hamiltonian}. Crucially, QGHD is a {\it quadratic theory}, so correlation functions can be calculated easily, at least numerically. [Higher-derivative and higher-order terms would lead to a generalization of the non-linear Luttinger liquid~\cite{imambekov2012one,imambekov2009universal} or nonlinear bosonization~\cite{abanov2005quantum,bettelheim2008quantum,stone2008classical,kulkarni2009nonlinear}; they are beyond the scope of this Letter.]

QGHD is the theory of a multi-component, spatially inhomogeneous, time-dependent, quantum fluctuating liquid with (locally) $q$ coupled components. Importantly, in the particular case of homogeneous time-independent split Fermi seas, we have checked~(see SM \cite{SM}) that it coincides with the multi-component quadratic Hamiltonian of Eli\"ens and Caux~\cite{eliens2016general,eliens2017quantum} (see also Refs.~\cite{fokkema2014split,vlijm2016correlations}). As noted by these authors, the case of a single component $q=1$ is nothing but the standard Luttinger liquid theory.

\vspace{0.1cm}
\noindent {\bf\em An example, and numerical check.}\; To illustrate the possibilities offered by QGHD, we consider the dynamics of the 1d Bose gas after a quench of the trapping potential from double-well, $V_0(x)= a_4 x^4 - a_2 x^2$, to harmonic, $V(x) = \omega^2 x^2/2$. The gas is initially in its ground state in $V_0(x)$, with a single pair of Fermi points (i.e. $q=1$) everywhere. At time $t>0$, after some fraction of the period of the trap $\tau = \frac{2\pi}{\omega}$, the contour $\Gamma_t$ gets deformed and a region appears near the boundaries with a split Fermi sea $q=2$. Hence this is a true out-of-equilibrium situation, not describable by standard hydrodynamics. This protocol mimics the famous quantum Newton's cradle~\cite{kinoshita2006quantum} and it can be realized experimentally (see e.g. Refs.~\cite{schemmer2019generalized,joseph2011observation}).

We focus on the equal-time density-density correlation function (Fig.~\ref{movie}). At a point $x$, the fluctuations of the particle density are measured by the operator 
\begin{equation}
	\delta \hat{\rho} (x,t) \, = \, \sum_s  \left| \frac{ds}{dx} \right| \delta \hat{\rho} (s) \, = \, \sum_{a}  \frac{1}{2\pi \hbar} \delta \hat{p}_a ,
\end{equation}
which is a sum over the $2q$ Fermi points at $(x,t)$. Its two-point function at time $t$ is
\begin{eqnarray} \label{rhorho}
	\left< \delta \hat{\rho} (x,t) \delta \hat{\rho} (x',t) \right>  &=&  \sum_{s} \sum_{s'} \left| \frac{ds}{dx} \right|  \left| \frac{ds'}{dx'} \right| G((s,t),(s',t)) , \nonumber\quad
\end{eqnarray}
where $ G((s,t),(s',t'))$ is the Green's function along the contour $G((s,t) ,(s' ,t')) \, = \, \left< \delta \hat{\rho} (s,t) \delta \hat{\rho} (s',t') \right>$. At $t=t'=0$, $G((s,0),(s',0))$ is the ground state correlation in the Hamiltonian $\hat{H}[\Gamma_0]$. At later times $G((s,t),(s',t'))$ satisfies the evolution equation derived from
\begin{equation}
	\frac{d}{dt} \delta \hat{\rho} (s,t) \, = \, \partial_s (v(s) \delta \hat{\rho} (s,t)) + \frac{i}{\hbar} [ \hat{H} [\Gamma_t], \delta \hat{\rho}(s,t) ] ,
\end{equation}
where $v(s) = v^{\rm eff} (\theta_a) \frac{dx}{ds}$ if $a$ labels the local Fermi point with parameter $s$. Importantly, $G((s,t),(s',t'))$ is of order $O(1)$ in the limit (\ref{eq:limit}), so we see that QGHD captures the first correction to the classical result (which is zero):
\begin{eqnarray}
	\frac{ \left< \delta \hat{\rho} (x,t) \delta \hat{\rho} (x',t) \right> }{\rho_{\rm cl.} (x) \rho_{\rm cl.} (x') 
} = O(\hbar^2) .
\end{eqnarray}

In Fig.~\ref{movie} we numerically evaluate the Green's function and compare the QGHD prediction (\ref{rhorho}) with a time-dependent Density-Matrix Renormalization Group (tDRMG)~\cite{dmrg-rev, itensor} simulation of the microscopic model. The dimensionless Lieb parameter $\gamma = \frac{\bar{g}}{\hbar \rho}$ is chosen to be of order $1$, so we are in the truly interacting regime of the 1d Bose gas, away from both the 
Gross-Pitaevski and the Tonks-Girardeau limits. 
The tDMRG simulation is performed for a lattice gas at very low density ($N \ll L$, where $L$ is the number of lattice sites)~\cite{schmidt2007exact,peotta2014quantum}, to be as close as possible to the continuum limit.
The largest number of particles accessible with this method is of order of $N \sim 20$~\cite{peotta2014quantum}, hence far from the thermodynamic limit. Consequently, finite-$N$ effects are large in our data, which we display for $N = 10, 20$ (and $L =800$). Still, the agreement between QGHD and numerics is good, and it improves as $\hbar$ decreases (i.e. $N\sim 1/\hbar$ increases). The tDMRG simulation becomes less accurate at large time; for this reason we stop the simulation at $t= 0.3 \tau$.
The limitations of tDMRG to small $N$ and small $t$ make the predictive power of QGHD even more apparent: QGHD does not suffer from those limitations as it works directly in the thermodynamic limit.

One interesting physical feature of Fig.~\ref{movie} is the divergence of the density-density correlation, in the thermodynamic limit, at the points where a change in the number of Fermi points occurs. They come from the Jacobians in Eq. (\ref{rhorho}) and are genuine predictions of the theory, valid for large enough $N$. The presence of these peaks can be explicitly confirmed by direct computations in the Tonks-Girardeau limit, where they are superimposed to Friedel oscillations~\cite{inpreparation} (see also Ref.~\cite{brun2018inhomogeneous} about the equilibrium case in a trap, where these divergences appear near the edges of the system), but they are a general consequence of QGHD at any interaction strength. At the small value $N=20$, the peaks' extent is smaller than that of Euler fluid cells, hence the peaks are washed away, as seen in the tDMRG result of Fig.~\ref{movie}.

\vspace{0.1cm}
\noindent {\bf\em Conclusion.}\; By focusing on the GHD description of the integrable 1d Bose gas in states of zero entropy, we showed that quantum effects which fall beyond the GHD description can be reconstructed by allowing quantum fluctuations of the Fermi contour.
We have been partially inspired by linear fluctuating hydrodynamics \cite{abanov2006hydrodynamics,spohn2014nlfh}, where
fluctuations are accessed by phenomenologically adding thermal noise to the linear response evolution of conserved fluid modes.
We follow the general principles of this theory, but instead of adding thermal noise, we use ideas from quantum fluids
(see e.g. \cite{abanov2006hydrodynamics}) in order to access quantum fluctuations.
To benchmark QGHD, we applied it to a zero entropy quench in the 1d Bose gas, providing exact predictions for the equal time density-density correlations, and checking that they are in good agreement with numerical tDMRG data obtainable for a small particle number and short times.

\vspace{0.1cm}
\begin{acknowledgments}
We thank S. Eli\"ens, M. Fagotti and J. de Nardis for discussions and A. Bastianello, E. Bettelheim, Y. Brun, A. De Luca, M. Collura, J. Viti and J.-M. St\'ephan for  collaboration on closely related topics. The DMRG simulation in Fig. 2 was done with iTensor~\cite{itensor}; we thank F. Pascale, J.-M. St\' ephan and T. Botzung for help with that simulation. We are grateful to the International Institute of Physics, Natal, Brazil, and to the University of Amsterdam, Netherlands, for hospitality during the completion of this work. Part of this work was supported by the CNRS ``D\'efi Infiniti'' MUSIQ (JD). PC and PR acknowledge support from ERC under Consolidator grant number 771536 (NEMO). BD acknowledges support from Royal Society under Leverhulme Trust Senior Research Fellowship SRF$\setminus$R1$\setminus$180103 ``Emergent hydrodynamics in integrable systems: non-equilibrium theory''; BD is also grateful to the Tokyo Institute of Technology, Tokyo, Japan, for funding and hospitality.
\end{acknowledgments}

\bibliography{qGHD}

\newpage

\begin{widetext}

\begin{center}
{\bf Supplementary material for ``Quantum Generalized Hydrodynamics''}
\end{center}

\section{Details of the derivation of Eq. (7) in the main text}

In this section we derive Eq. (7) in the main text. We start from the GHD equations (4a)-(4b) in the main text. Parametrizing the
contour locally as $\theta_a (x,t)$ and injecting this parametrization into Eq.~(4a), one gets
\begin{eqnarray*}
	\frac{d}{dt} \left( \begin{array}{c}
		x_t(s) \\
		\theta_a (x_t(s), t)
	\end{array} \right) &=& \left(  \begin{array}{c}
		v^{\rm eff} (x_t(s), \theta_a(x_t(s),t) ) \\
		a^{\rm eff} (x_t(s), \theta_a(x_t(s),t) )
	\end{array} \right)  .
\end{eqnarray*}
The second line reads $\partial_t \theta_a (x_t(s), t) + (\partial_t x_t) \partial_x \theta_a (x_t(s), t)  =  a^{\rm eff} (x_t(s), t)$. 
Then, plugging the first line $\partial_t x_t = v^{\rm eff}$ into it, one gets the zero-entropy GHD equation of Ref.~\cite{doyon2017large}:
\begin{equation}
	 \partial_t \theta_a (x, t) + v^{\rm eff}(x,\theta_a)  \partial_x \theta_a (x, t)  = a^{\rm eff} (x,t).
\end{equation}
Finally, we use Eq. (6) in the main text: $\partial_t p  + \partial_x \epsilon + a^{\rm eff} 1^{\rm dr} = 0$. With $p_a = p(\theta_a)$ and $\epsilon_a = \epsilon(\theta_a)$, this gives
\begin{eqnarray*}
	\partial_t p_a + \partial_x \epsilon_a &=&  [ \partial_t p  + \partial_x \epsilon]  (\theta_a) + (\partial_t \theta_a) (\partial_{\theta} p)  + (\partial_x \theta_a)  (\partial_{\theta} \epsilon)  \\
	&=& - a^{\rm eff} (\theta_a) 1^{\rm dr}  (\theta_a)  + (\partial_{\theta} p) \left[ \partial_t \theta_a +  \frac{\partial_{\theta} \epsilon}{\partial_{\theta} p}  \partial_x \theta_a\right] \\
	&=& - a^{\rm eff} (\theta_a) 1^{\rm dr}  (\theta_a)  + (\partial_{\theta} p) \left[ \partial_t \theta_a + v^{\rm eff} (\theta_a) \partial_x \theta_a  \right] \\
	&=&  \left(  - 1^{\rm dr}  (\theta_a)  + \partial_{\theta} p \right) a^{\rm eff} (\theta_a)  .
\end{eqnarray*}
Finally, using the identity $\partial_{\theta} p = ({\rm id}')^{\rm dr} (\theta) = 1^{\rm dr} (\theta) $ ---see Eqs. (\ref{sm:eq:drDr}) and (\ref{sm:eq:k}) in this Supplemental Material---, the last line cancels and one gets $\partial_t p_a + \partial_x \epsilon_a = 0$, which is Eq. (7) in the main text.

\section{Onsager reciprocity relation and consistency with Eli\"ens-Caux formalism}

In this section we expose in full details the Thermodynamic Bethe Ansatz (TBA) calculations that are useful to arrive at the Onsager reciprocity relation (Eq. (10) in the main text) and at the consistency of our results with the ones previously obtained by Eli\"ens and Caux \cite{eliens2016general,eliens2017quantum}.

\subsection{Useful definitions: shift function, ``dressing'' and ``Dressing''}

The shift function is defined as follows (see e.g. chapter 1 of Ref.~\cite{korepin1997quantum}). Adding a (fermionic) particle with rapidity $\theta$ results in a global shift of all rapidities measured by the shift function
\begin{eqnarray}
	\label{sm:eq:shift}	
F(\theta | \theta') &=& \frac{\phi(\theta-\theta')}{2\pi} + \int_M \frac{d\lambda}{2\pi} \varphi(\theta-\lambda) F(\lambda | \theta') 
=	\frac{[\phi(.-\theta')]^{\rm dr} (\theta)}{2\pi} ,
\end{eqnarray}
where $M = [\theta_1,\theta_2] \cup [\theta_3,\theta_4] \cup \dots \cup [\theta_{2n-1},\theta_{2n}]$ is the split Fermi sea. The dressing is the linear operation $f(\theta) \mapsto f^{\rm dr}(\theta)$ defined by the integral equation
\begin{equation}
	\label{sm:eq:dr}
	f^{\rm dr} (\theta) = f(\theta) + \int_M \frac{d\theta'}{2\pi} \varphi(\theta-\theta') f^{\rm dr}(\theta') .
\end{equation}
A very useful property of this dressing operation is that it is symmetric,
\begin{equation}
	\int_M \frac{d\theta}{2\pi} f^{\rm dr} (\theta) g(\theta) \, = \, \int_M \frac{d\theta}{2\pi} f (\theta) g^{\rm dr}(\theta) .
\end{equation}
This is not the ``physical'' dressing though. The physical dressing (or ``Dressing'') is rather defined as
\begin{equation}
	\label{sm:eq:Dressing}
	f^{\rm Dr} (\theta) \, = \, f(\theta) - \int_M d \theta' \, f'(\theta')  \, F(\theta' | \theta) .
\end{equation}
The two kinds of dressing are related as follows:
\begin{eqnarray}
	\label{sm:eq:drDr}
\nonumber	( f^{\rm Dr} )' (\theta) & = & f'(\theta) - \partial_\theta \left[ \int_M \frac{d\theta'}{2\pi} f'(\theta')  [\phi(.- \theta)]^{\rm dr}(\theta') \right] 
=f'(\theta)+  \partial_\theta \left[ \int_M \frac{d\theta'}{2\pi} (f')^{\rm dr}(\theta')  \phi( \theta- \theta') \right] \\
& = & f'(\theta)  + \int_M \frac{d\theta'}{2\pi}  \varphi( \theta- \theta') (f')^{\rm dr}(\theta')   
=	(f')^{\rm dr}(\theta) .
\end{eqnarray}
In particular, adding a (fermionic) excitation with rapidity $\theta$ to a state results in a change of the total momentum and energy by an amount
\begin{eqnarray}
	\label{sm:eq:k}
	p  &=&  {\rm id}^{\rm Dr} (\theta) \, = \, \theta - \int_M d\theta' \, F(\theta' | \theta) ,\\
	\epsilon  &=&  E^{\rm Dr} (\theta) \, = \, \theta - \int_M d\theta' \, E'(\theta')  F(\theta' | \theta) \nonumber.
\end{eqnarray}
This is what we use in the main text, in the discussion which precedes Eq. (6) there.

\subsection{The matrix $F$ of shifts at the Fermi points}\label{SectFs}

Differentiating the definition of the dressing (\ref{sm:eq:dr}) w.r.t $\theta$ gives
\begin{equation}
	\label{sm:eq:fdrprime}
	(f')^{\rm dr} (\theta) = (f^{\rm dr})'(\theta) + \sum_c \frac{\sigma_c}{2\pi} f^{\rm dr} (\theta_c) [\varphi(.-\theta_c) ]^{\rm dr} (\theta) .
\end{equation}
Then using the definition of the dressing, the antisymmetry of $\phi$, and the above formula with $f(.) = \phi(.-\theta')$, one gets (see also formula (7.26) in Ref.~\cite{eliens2017quantum})
\begin{eqnarray*}
	 F(\theta | \theta') + F(\theta' | \theta)  &=&
	\int_M \frac{d\lambda}{2\pi} \left[ \varphi(\theta-\lambda) \frac{[\phi(.-\theta')]^{\rm dr}(\lambda)}{2\pi} + \varphi(\theta'-\lambda) \frac{[\phi(.-\theta)]^{\rm dr}(\lambda)}{2\pi} \right] \\
	&=& \int_M \frac{d\lambda}{2\pi} \left[ \varphi(\lambda-\theta) \frac{[\phi(.-\theta')]^{\rm dr}(\lambda)}{2\pi} + \varphi(\lambda-\theta') \frac{[\phi(.-\theta)]^{\rm dr}(\lambda)}{2\pi} \right] \\
	&=& \int_M \frac{d\lambda}{2\pi} \left[ [\varphi(.-\theta)]^{\rm dr} (\lambda) \frac{\phi(\lambda-\theta')}{2\pi} + \varphi(\lambda-\theta') \frac{[\phi(.-\theta)]^{\rm dr}(\lambda)}{2\pi} \right]  \\
	&=& \int_M \frac{d\lambda}{2\pi} \left[ ([\phi(.-\theta)]^{\rm dr} )'(\lambda) \frac{\phi(\lambda-\theta')}{2\pi} + \frac{[\phi(.-\theta)]^{\rm dr}(\lambda)}{2\pi} \phi'(\lambda-\theta')  \right]  \\
	&& +  \int_M \frac{d\lambda}{2\pi} \left[ \sum_c \frac{\sigma_c}{2\pi} [\phi(.-\theta)]^{\rm dr} (\theta_c) [\varphi(.- \theta_c)]^{\rm dr}(\lambda)  \frac{\phi(\lambda-\theta')}{2\pi}  \right] \\
	&=& \int_M \frac{d\lambda}{2\pi} \frac{\partial}{\partial \lambda}\left[  [\phi(.-\theta)]^{\rm dr} (\lambda) \frac{\phi(\lambda-\theta')}{2\pi} \right]  \\
	&& +  \sum_c \frac{\sigma_c}{2\pi} [\phi(.-\theta)]^{\rm dr} (\theta_c) \int_M \frac{d\lambda}{2\pi}  \varphi(\lambda- \theta_c)  \frac{[\phi(.-\theta')]^{\rm dr}(\lambda)}{2\pi}  \\	
	&=& \sum_a \frac{\sigma_a}{2\pi}   [\phi(.-\theta)]^{\rm dr} (\theta_a) \frac{\phi(\theta_a-\theta')}{2\pi}    \\
	&& +  \sum_c \frac{\sigma_c}{2\pi} [\phi(.-\theta)]^{\rm dr} (\theta_c) \left(  \frac{[\phi(.-\theta')]^{\rm dr}(\theta_c)}{2\pi} -  \frac{\phi(\theta_c-\theta')}{2\pi} \right) \\	
	&=&  \sum_a  F(\theta_a | \theta) \sigma_a  F(\theta_a | \theta').
\end{eqnarray*}
In particular, if we define the $2n \times 2n$ matrix $F$ as
\begin{equation}
	F_{ab} \, := \, F(\theta_a | \theta_b), 
\end{equation}
then the following identity holds:
\begin{equation}
	\label{sm:eq:FFdag}
	F + F^\dagger \, = \, F^\dagger  \sigma  F.
\end{equation}

\subsection{The Eli\"ens-Caux matrix}

A key object in the papers of Eli\"ens and Caux is the following $2q \times 2q$ matrix $M$ (see formulas (7.51) and (7.52) in Ref.~\cite{eliens2017quantum}), defined in terms of the Jacobian of the transformation from the Fermi rapidities $\{ \theta_a\}_{1 \leq a \leq 2q}$ to the Fermi momenta $\{ p_a\}_{1 \leq a \leq 2q}$:
\begin{equation}
	\label{sm:eq:EC}
	M_{ab} \, = \, \frac{1}{1^{\rm dr}(\theta_b)} \frac{\partial p_a}{ \partial \theta_b} .
\end{equation}
We call it the {\it Eli\"ens-Caux matrix}. It can be expressed in terms of the above matrix $F$:
\begin{equation}
	\label{sm:eq:VF}
	M \, = \, 1 - F^\dagger  \sigma \qquad \quad ({\rm in \; components}, \;M_{ab} = \delta_{ab} - \sigma_b F(\theta_b| \theta_a)) .
\end{equation}
This is obtained as follows. Differentiating the definition of the shift function (\ref{sm:eq:shift}), and using the definition of the dressing, one gets
\begin{eqnarray}
	\delta F(\theta | \theta') &=& \sum_b \frac{\sigma_b}{2\pi} [\varphi(. - \theta_b)]^{\rm dr}(\theta) \, F(\theta_b | \theta') \, \delta \theta_b .
\end{eqnarray}
This leads to the variation of the ``Dressed'' function (\ref{sm:eq:Dressing}),
\begin{eqnarray}
	\nonumber \delta f^{\rm Dr} (\theta ) &=& - \sum_b \sigma_b \, f'(\theta_b) \,  F(\theta_b | \theta) \delta \theta_b - \int_M d \theta' \, f'(\theta') \, \delta F(\theta' | \theta) \\
	\nonumber &=&  - \sum_b \sigma_b \, f'(\theta_b) \,  F(\theta_b | \theta) \delta \theta_b - \sum_b \frac{\sigma_b}{2\pi}  \left( \int_M d\theta' \, f'(\theta') \,[\varphi(. - \theta_b)]^{\rm dr}(\theta') \right) \, F(\theta_b | \theta) \, \delta \theta_b  \\
	\nonumber &=&  - \sum_b \sigma_b \, f'(\theta_b) \,  F(\theta_b | \theta) \delta \theta_b - \sum_b \sigma_b  \left( \int_M \frac{d\theta'}{2\pi} \, (f')^{\rm dr}(\theta') \,\varphi(\theta' - \theta_b) \right) \, F(\theta_b | \theta) \, \delta \theta_b  \\
	\nonumber &=&  - \sum_b \sigma_b \, f'(\theta_b) \,  F(\theta_b | \theta) \delta \theta_b - \sum_b \sigma_b  \left(  (f')^{\rm dr}(\theta_b) - f'(\theta_b) \right) \, F(\theta_b | \theta) \, \delta \theta_b  \\
	&=&- \sum_b \sigma_b \, F(\theta_b | \theta) \,  (f')^{\rm dr}(\theta_b) \, \delta \theta_b .
\end{eqnarray}
Consequently, we have the identity
\begin{eqnarray}
	\label{sm:eq:deltafDr}
\nonumber	\delta ( f^{\rm Dr} (\theta_a)) &=&   ( f^{\rm Dr} )' (\theta_a)  \delta \theta_a \, + \,   ( \delta f^{\rm Dr}  ) (\theta_a)  
=( f' )^{\rm dr} (\theta_a)  \delta \theta_a \, - \, \sum_b \sigma_b \, F(\theta_b | \theta_a) \,  (f')^{\rm dr}(\theta_b)  \, \delta \theta_b \\
	&=&  \sum_b  \left( \delta_{ab} - \sigma_b \, F(\theta_b | \theta_a)  \right)  (f')^{\rm dr} (\theta_b)  \delta \theta_b    . 
\end{eqnarray}
In particular, plugging the definition of $p$ (Eq. (\ref{sm:eq:k})) into that formula, one gets
\begin{equation*}
	\delta p_a \, = \, \delta ( {\rm id}^{\rm Dr} (\theta_a) ) 
	\, =\,  \sum_b  \left( \delta_{ab} - \sigma_b \, F(\theta_b | \theta_a)  \right)  1^{\rm dr} (\theta_b)  \delta \theta_b  \, ,
\end{equation*}
which is the Eli\"ens-Caux relation (\ref{sm:eq:VF}). Notice that, more generally, the Eli\"ens-Caux matrix appears in the derivative of $f^{\rm Dr} (\theta_a)$ for any function $f$, as a consequence of (\ref{sm:eq:deltafDr}):
\begin{equation}
	\label{sm:eq:dfDR}
		\frac{\partial ( f^{\rm Dr} (\theta_a) )}{\partial \theta_b} \, = \,  M_{ab}  \, (f')^{\rm dr} (\theta_b) .
\end{equation}

\subsection{Symplecticity of the Eli\"ens-Caux matrix}

A particularly remarkable property of the Eli\"ens-Caux matrix is that it satisfies
\begin{equation}
	\label{sm:eq:symplecticity}
	M^{-1} = \sigma M^\dagger \sigma .
\end{equation}
This follows from Eqs. (\ref{sm:eq:VF}) and (\ref{sm:eq:FFdag}): $M \sigma M^\dagger \sigma = (1-F^\dagger \sigma) \sigma (1 - \sigma F) \sigma = 1 - ( F + F^\dagger - F^\dagger \sigma F) \sigma = 1$, so $M$ is invertible and its inverse is $\sigma M^\dagger \sigma$.

\subsection{The Eli\"ens-Caux matrix and the flux Jacobian $\mathsf A$}

In the main text, a central role is played by the flux Jacobian, defined as
\begin{equation}\label{sm:eq:A_EC}
	\mathsf A_{a}^{~b} = \frac{\partial E^{\rm Dr} (\theta_a)}{\partial p_b}  .
\end{equation}
This flux Jacobian can also be expressed in terms of the Eli\"ens-Caux matrix, using formula (\ref{sm:eq:dfDR}):
\begin{eqnarray}
\nonumber	\mathsf A_{a}^{~b}  &=& \frac{ \partial }{\partial p_b} E^{\rm Dr}(\theta_a) 
=  \sum_c \frac{ \partial \theta_c }{\partial p_b}  \frac{\partial ( E^{\rm Dr}(\theta_a) ) }{\partial \theta_c} 
=  \sum_c    (1^{\rm dr}(\theta_c))^{-1} [M^{-1}]_{cb}  \frac{\partial (E^{\rm Dr}(\theta_a))}{\partial \theta_c}   \\
	&=&  \sum_c    (1^{\rm dr}(\theta_c))^{-1} [M^{-1}]_{cb}  M_{ac}  (E')^{\rm dr} (\theta_c) 
	= [M v^{\rm eff} M^{-1}]_{ab}.
\end{eqnarray}
In the last line we have used the definition of the effective velocity, $v^{\rm eff}(\theta_c) = \frac{(E')^{\rm dr} (\theta_c)}{ 1^{\rm dr} (\theta_c) }$, see formula (4b) in the main text.

In other words, the flux Jacobian is diagonalized by the Eli\"ens-Caux matrix. The identity (\ref{sm:eq:A_EC}) is the key to derive both the Onsager reciprocity relation and to check the consistency of our Hamiltonian with the one of Eli\"ens and Caux.

\subsection{Onsager reciprocity relation}

Defining $\mathsf A^{ab} = \sigma_b \mathsf A_a^{~b}$ as in the main text, we must show that $\mathsf A^{ab}  = \mathsf A^{ba}$. Equivalently, using formula (\ref{sm:eq:A_EC}), we must show that
\begin{equation}
	\sigma  M v^{\rm eff} M^{-1} \, = \, [ \sigma M v^{\rm eff} M^{-1} ]^\dagger .
\end{equation}
In that form, the reciprocity relation is a straightforward consequence of the symplecticity of $M$ (formula (\ref{sm:eq:symplecticity})), and of the fact that $\sigma$ and $v^{\rm eff}$ are diagonal matrices.

\subsection{Consistency of the QGHD Hamiltonian with the one of Eli\"ens and Caux}

In their study of homogeneous, time-independent split Fermi seas, Eli\"ens and Caux write the following multi-component Luttinger liquid Hamiltonian (see Eqs. (6.17) and (6.18) in Ref. \cite{eliens2017quantum}):
\begin{equation}
	H_{EC} \, = \, \sum_{a=1}^{2q} \frac{\sigma_a v^{\rm eff}_a}{2\pi} \int dx (\partial_x \hat{\varphi}_{a})^2 ,
\end{equation}
where the $\hat{\varphi}_a (x,t)$ are $2q$ independent chiral bosonic modes, which are related to our operators $\delta \hat{p}_a$ (see the main text for definition) as
\begin{equation}
	\delta \hat{p}_a = \sqrt{2 \hbar} \sum_{b=1}^{2q} M_{ab} \hat{\varphi}_b (x,t) .
\end{equation}
Thus, their Hamiltonian may be written as
\begin{equation}
	H_{EC} \, = \, \frac{1}{4\pi \hbar} \int dx \sum_{a,b} \delta \hat{p}_a [(M^{-1})^{\dagger} \sigma v^{\rm eff} M^{-1}]_{ab} \delta \hat{p}_b .
\end{equation}
Using again the symplecticity of $M$ and formula (\ref{sm:eq:A_EC}), we see that this is identical to our QGHD Hamiltonian, see Eq. (12) in the main text.

We emphasize that, although our QGHD Hamiltonian is identical to the one of Eli\"ens and Caux for homogeneous, time-independent split Fermi seas, our QGHD formalism is a non-trivial extension of their results. This is because, when the parameters in the Hamiltonian become position- and time-dependent, there are in principle many different
terms involving derivatives of the classical GHD solution $\{ \theta_a (x,t) \}$ which could enter, which would result in a Hamiltonian different from ours, yet which would still coincide with the one of Eli\"ens and Caux in the case where the derivatives vanish. For a more thorough discussion in the spatially inhomogeneous static case, see Ref.~\cite{brun2018inhomogeneous}.

\section{QGHD for more general integrable models}

In the main text, the QGHD theory was developed for the Lieb-Liniger model. This is a Galilean invariant model with a single, fermionic quasiparticle specie, and with a specific scattering phase. Here we generalise the setup to Bethe-ansatz integrable models ---quantum field theories or quantum chains--- with arbitrary scattering and an arbitrary number of species, with the sole condition that all TBA quasiparticle species be of fermionic statistics. This includes, for instance, the XXZ quantum spin chain and the sine-Gordon model.

One of the strengths of GHD is that its general structure stays valid for a very wide family of integrable models. The main ingredients are a {\em spectral space}, the space of quasiparticle species and their allowed momenta, a {\em scattering phase function}, the logarithm of the (TBA-diagonalised) scattering matrix, and a {\em statistical function}, essentially the form of the filling function as fixed by the statistics, entering for instance the TBA expression of the thermodynamic entropy. See \cite{notes} for how these ingredients are used in GHD. Bethe-ansatz integrable models present a variety of structures for their eigenstates, but in many important cases, in the thermodynamic limit, GHD can be brought to this normal form, with these ingredients. For instance, even if the microscopic model has a single particle specie, often particular Bethe-ansatz solutions, such Bethe roots organising themselves into strings in the complex plane, are identified with new TBA quasiparticle species; this happens in the XXZ model (see \cite{bertini2016transport} for its GHD). Also, if the bare Bethe-ansatz scattering matrix is not diagonal, then the internal structure can be diagonalised (nested Bethe ansatz) and a new set of emergent, diagonally-scattering quasiparticles appear in the TBA; this happens in the sine-Gordon model and for the Yang-Gaudin gas and Hubbard model (see \cite{PhysRevB.100.035108,mestyan2019GHD,PhysRevB.96.081118,nozawa2020generalized} for their GHD). In all these examples, the emergent TBA quasiparticles come in many species, with well-defined scattering phase function and fermionic statistics, and the general framework of GHD applies.

In the general form of GHD, as compared to the main text, we therefore make the changes:
\begin{eqnarray}
	\theta &\longrightarrow&  (\theta,u)\quad \mbox{(spectral space)}\\
	\phi(\theta-\theta') &\longrightarrow&
	\phi((\theta,u),(\theta',u'))\quad \mbox{(scattering phase)}
\end{eqnarray}
where $u,u'$ run over the quasiparticle species. Further, we make the replacement
\begin{equation}
	\int d\theta \longrightarrow  \sum_u\int d\theta.
\end{equation}
In general, one takes $\theta$s as ``rapidities", which  parametrise the bare momentum of the quasiparticle as
\begin{equation}
	P(\theta,u).
\end{equation}
The rapidity parametrisation of the momentum is a choice, and with a good choice, one often has
\begin{equation}
	\phi((\theta,u),(\theta',u')) = \phi(\theta-\theta',u,u').
\end{equation}
Since such a choice is possible in many important models, for simplicity we assume it below (but this is not essential). Further, the unitarity of the scattering matrix in parity-invariant models takes the simple form
\begin{equation}\label{symuu}
	\phi(\theta-\theta',u,u') = -\phi(\theta'-\theta,u',u),\quad
	\varphi(\theta-\theta',u,u') = \varphi(\theta'-\theta,u',u)
\end{equation}
with the differential scattering phase $\varphi(\theta-\theta',u,u') = \partial_\theta \phi(\theta-\theta',u,u')$. We also assume this. Finally, the momentum parametrisation may have either everywhere positive, or everywhere negative, $\theta$-derivative, $P'(\theta,u)>0\;\forall\;\theta$ or $P'(\theta,u)<0\;\forall\;\theta$. This corresponds to its (in general specie-dependent) parity $\sigma_u=\pm1$. In general, the parity factor $\sigma_u$ must be inserted at various places, in a way that is fully determined by the transformation property of the mathematical objects involved (for instance, the integration measure is $d\theta\,\sigma_u$). See the explanations in \cite{10.21468/SciPostPhys.6.4.049}. Below we assume for lightness of notation that, as in the Lieb-Liniger model, $\sigma_u=+1$.

In the general setting, the occupation function $n(\theta,u)$ still exists, and diagonalises the GHD equation (these are the fluid's normal modes). With Fermi-sea fillings, we have Fermi points $\theta_a(u)$ which now depend on the quasiparticle specie $u$; the range of $a\in\{1,\ldots,2q_u\}$ also depends on $u$. With this additional dependence, Equation (4a) in the main text stays valid. Equation (4b) is modified, in general, to
\begin{equation}
	v^{\rm eff}  \, = \, (\partial_\theta E)^{\rm dr} / (\partial_\theta P)^{\rm dr}   ,\quad
	a^{\rm eff}  \, = \, -(\partial_xE)^{\rm dr} / (\partial_\theta P)^{\rm dr}.
\end{equation}
With many quasiparticle species, Equation (5) of the main text in general does not hold (but it was not used in any of the derivations).

Using the physical momentum and energy of an excitation as
\begin{eqnarray}
p(\theta,u) &=& P(\theta,u) + \sum_{u'}\int \frac{d\theta'}{2\pi} \phi(\theta-\theta',u,u') n(\theta',u')(\partial_\theta P)^{\rm dr}(\theta',u')\\
\epsilon(\theta,u) &=&  E(x,\theta,u) + \sum_{u'}\int \frac{d\theta'}{2\pi} \phi(\theta-\theta',u,u')n(\theta',u')(\partial_\theta P)^{\rm dr}(\theta',u')v^{\rm eff}(\theta',u')
\end{eqnarray}
it still holds that $\partial_\theta \epsilon / \partial_\theta p = v^{\rm eff}$, as in fact
\begin{equation}
	\partial_\theta p = (\partial_\theta P)^{\rm dr},\quad
	\partial_\theta \epsilon = (\partial_\theta E)^{\rm dr}.
\end{equation}
The conservation Equation (6) (main text) becomes
\begin{equation}
	\partial_t p + \partial_x \epsilon + a^{\rm eff} (\partial_\theta P)^{\rm dr}=0.
\end{equation}
Defining $p_{a,u} = p(\theta_a,u)$  and  $\epsilon_{a,u} = \epsilon(\theta_a,u)$, Eqs.~(7) and (8) (main text) become
\begin{equation}
	\partial_t p_{a,u} + \partial_x \epsilon_{a,u} = 0
\end{equation}
and
\begin{equation}
	\partial_t \delta p_{a,u} + \sum_{u'}\sum_{b=1}^{2q_{u'}}\partial_x[\mathsf A_{a,u}^{~b,u'} \delta p_{b,u'}] = 0,
\end{equation}
where $ \mathsf A_{a,u}^{~b,u'} = \partial \epsilon_{a,u}/\partial p_{b,u'}$. 

Finally, we can work out the identities related to the Eli\"ens-Caux matrix in the same way. Defining
\begin{equation}
	F_{(a,u),(b,u')} \,:=\, F(\theta_a,u | \theta_b,u')
\end{equation}
the derivation in Section II.B is essentially unchanged, where we use the symmetry \eqref{symuu}, and \eqref{sm:eq:FFdag} stays true, with $\sigma_{a,u} = (-1)^a$. For Section C, we have, instead of \eqref{sm:eq:EC},
\begin{equation}
	M_{(a,u),(b,u')} \, = \, \frac{1}{(\partial_\theta P)^{\rm dr}(\theta_b,u')} \frac{\partial p_{a,u}}{ \partial \theta_{b,u'}} .
\end{equation}
Therefore, we still have
\begin{equation}
	M = 1-F^\dag \sigma
\end{equation}
and the important relation \eqref{sm:eq:symplecticity} remains valid, as well as
\begin{equation}
	\mathsf A_{a,u}^{~b,u'} = [Mv^{\rm eff}M^{-1}]_{(a,u),(b,u')}.
\end{equation}
Thus the Onsager reciprocity relation still holds.

The main results, Equations (11b) and (12) in the main text, are therefore valid in the multi-specie case, with the replacement of single-indices for the Fermi-sea boundaries, by double-indices for the additional information of the quasiparticle specie,
\begin{equation}
	a \longrightarrow (a,u).
\end{equation}

Finally, we mention an important conceptual point. It is clear that the technical derivation above is entirely insensitive to the statistics of the quasiparticles: the only requirement is to start with an occupation function that is of (multiple-)Fermi-sea type. Although the family of such occupation functions is most natural with fermionic statistics, it is {\em still allowed}, and is invariant under time evolution, with other statistics as well. However, {\em it is only with fermionic statistics that such occupation functions correspond to zero-entropy states}, where quantum fluctuations are expected to provide the leading correlations. Despite the formal validity of the technical derivation independently of the statistics, it is only with fermionic statistics that we expect the quantum hamiltonian (12) (main text) to correctly describe correlations (but this is just a minor restriction since all known interacting integrable models have fermionic quasiparticles).


\end{widetext}

\end{document}